\documentclass[preprint]{aastex}
\usepackage{graphicx}

\shorttitle{MIRIM, The Imager of the MIRI instrument}
\shortauthors{Bouchet et al.}

\begin{document}


\title{The Mid-Infrared Instrument for the James Webb Space Telescope, III: MIRIM, The MIRI Imager}

\author{Patrice Bouchet\altaffilmark{1}, Macarena Garc\'ia-Mar\'in\altaffilmark{2}, P.-O. Lagage\altaffilmark{1}, J\'erome Amiaux\altaffilmark{1}, 
J.-L. Augu\'eres\altaffilmark{1}, Eva Bauwens\altaffilmark{3}, J. A. D. L. Blommaert\altaffilmark{3,4,5}, C. H. Chen\altaffilmark{6},
  {\"O}.H. Detre\altaffilmark{7}, Dan Dicken\altaffilmark{1}, D. Dubreuil\altaffilmark{1}, Ph. Galdemard\altaffilmark{8}, R. Gastaud\altaffilmark{8}, A. 
Glasse\altaffilmark{9}, K. D. Gordon\altaffilmark{6,10}; F. Gougnaud\altaffilmark{8} ; Phillippe Guillard\altaffilmark{11}; K. Justtanont\altaffilmark{12}, Oliver 
Krause\altaffilmark{7}, Didier Leboeuf\altaffilmark{8}, Yuying Longval\altaffilmark{11}, Laurant Martin\altaffilmark{13}, Emmanuel Mazy\altaffilmark{14}, Vincent Moreau\altaffilmark{1}, 
G\"oran Olofsson\altaffilmark{12}, T. P. Ray\altaffilmark{15}, J.-M. Reess\altaffilmark{16}, Etienne Renotte\altaffilmark{14}, M. E.  Ressler\altaffilmark{17}, Samuel 
Ronayette\altaffilmark{1}, Sophie Salasca\altaffilmark{18}, Silvia Scheithauer\altaffilmark{7}, Jon Sykes\altaffilmark{19}, M. P.  Thelen\altaffilmark{17}, Martyn 
Wells\altaffilmark{9}, David Wright\altaffilmark{20}, G. S. Wright\altaffilmark{9}}

\altaffiltext{1}{Laboratoire AIM Paris-Saclay, CEA-IRFU/SAp, CNRS, Universit\'e Paris Diderot, F-91191 Gif-sur-Yvette, France}
\altaffiltext{2}{I. Physikalisches Institut, Universit\"at zu K\"oln, Z\"ulpicher Str. 77,  50937 K\"oln, Germany }
\altaffiltext{3}{Institute of Astronomy KU Leuven, Celestijnenlaan 200D,3001 Leuven, Belgium}
\altaffiltext{4}{Department of Physics and Astrophysics, Vrije Universiteit Brussel, Belgium}
\altaffiltext{5}{Flemish Institute for Technological Research (VITO), Boeretang 200,2400 Mol, Belgium}
\altaffiltext{6}{Space Telescope Science Institute, 3700 San Martin   Drive, Baltimore, MD, 21218, USA}
\altaffiltext{7}{Max Planck Institute f\"ur Astronomy (MPIA), K\"onigstuhl 17, D-69117 Heidelberg, Germany}
\altaffiltext{8}{DSM/Irfu/SIS, CEA-Saclay, F-91191 Gif-sur-Yvette, France}
\altaffiltext{9}{UK Astronomy Technology Centre, STFC, Royal Observatory Edinburgh, Blackford Hill, Edinburgh EH9 3HJ, UK}
\altaffiltext{10}{Sterrenkundig Observatorium, Universiteit Gent, Gent, Belgium}
\altaffiltext{11}{Institut d’Astrophysique Spatiale, CNRS (UMR 8617), Université Paris-Sud 11, Batiment 121, Orsay, France}
\altaffiltext{12}{Chalmers University of Technology, Onsala Space Observatory, S-439 92 Onsala, Sweden}
\altaffiltext{13}{Laboratoire d'Astrophysique de Marseille (LAM), Université d'Aix Marseille, CNRS (UMR 7326), 13388 Marseille, France}
\altaffiltext{14}{Centre Spatial De Li\`ege, Avenue du Pre Aily, B-4031, Angleur, Belgium}
\altaffiltext{15}{Dublin Institute for Advanced Studies, School of Cosmic Physics, 31 Fitzwilliam Place, Dublin 2, Ireland}
\altaffiltext{16}{LESIA, Observatoire de Paris-Meudon, CNRS, Universit\'e Pierre et Marie Curie, Universit\'e Paris Diderot, 5 Place Jules Janssen, F-92195 Meudon, France}
\altaffiltext{17}{Jet Propulsion Laboratory, California Institute of Technology, 4800 Oak Grove Dr. Pasadena, CA 91109, USA}
\altaffiltext{18}{CEA-IRFM/SIPP/GIPM, Centre d'Etudes de Cadarache, F-13108 Saint-Paul-lez-Durance, France}
\altaffiltext{19}{Department of Physics and Astronomy, Univ. of Leicester, University Road, Leicester, LE1 7RH, UK}
\altaffiltext{20}{Stinger Ghaffarian Technologies, Inc., Greenbelt, MD, USA}
\altaffiltext{}{}
\altaffiltext{}{}
\altaffiltext{}{}

\email{Patrice.Bouchet@cea.fr}






\begin{abstract}

In this article, we describe the MIRI Imager module (MIRIM), which provides 
broad-band imaging in the 5 -- 27 $\mu$m wavelength range for the 
James Webb Space Telescope. The imager has a 0\farcs11 pixel scale 
and a total unobstructed view of 74$''$ $\times$ 113$''$.  
The remainder of its nominal 113$''$ $\times$ 113$''$ field is occupied by the coronagraphs and the low resolution spectrometer. 
We present the instrument optical and mechanical  design. We show that the test data, as measured during the test campaigns undertaken at CEA-Saclay, at the Rutherford Appleton Laboratory, and at the NASA Goddard Space Flight Center, indicate that the instrument complies with its design requirements and goals. We also discuss the operational requirements (multiple dithers and exposures) needed for optimal scientific utilization of the MIRIM.  

\end{abstract}



\section{Introduction}

The MIRI instrument concept and design have been extensively described in \citet{Renouf06}, as well as Wright et al. (2014: hereafter Paper II, in this volume). The first test results achieved at CEA-Saclay have been reported in \citet{Ami08}.
We give here a brief observer-oriented description of the Imager module of MIRI, MIRIM, together with a summary of its performance as measured from the test runs that took place at CEA-Saclay in 2007 and 2008 (for the optical quality), at the Rutherford Appleton Laboratory (RAL, UK) where the Flight Model (FM) has been extensively tested
in 2011, and at the NASA Goddard Space Flight Center (GSFC, USA) where the instrument was mounted in the Integrated Science Instrument Module (ISIM) of the JWST in October 2013. 

The MIRIM provides broad-band imaging, phase-mask coronagraphy, Lyot
coronagraphy, and low-resolution slit and slitless spectroscopy (LRS) using a single $1024 \times 1024$ pixel 
Si:As IBC sensor chip assembly (SCA), with 25 $\mu$m pixels. Kendrew et al. (2014, hereafter Paper IV in this volume), and Boccaletti et al. (2014, Paper V), provide the full description of the LRS and the coronagraphs, respectively. 
As shown in Figure 1, the imager has a pixel scale of 0\farcs11 and a total field of view of $113''  \times 113''$; 
however, the field of view of its clear aperture for traditional imaging is $74'' \times113''$ because the coronagraph masks and 
the LRS are fixed on one side of the focal plane and occupy an area of $39'' \times113''$. 
The only moving part in the imager is an 18 position filter wheel (see details in Paper II)). 

This paper gives an overview of the MIRIM optics (\S 2) and the mechanical design (\S 3), followed by a comparison of the design with test measurements (\S 4). It 
then moves to describing the use of the instrument, including operations (\S 5), and calibration (\S 6).

\section{MIRIM Optics}

\subsection{The Optical Design}

Paper II discusses the overall design and construction of the MIRI. The manufacture and assembly of the instrument
are described by \citet{Ami08}. Here, we give a more detailed description of the imager optical and mechanical layout and how it interfaces with the rest of the instrument. 

Both the imager and spectrometer channels are fed by a single common pick-off mirror (POM) placed in front of the JWST Optical Telescope Assembly (OTA) focal plane. The light then passes into the MIRI Input Optics and Calibration (IOC) unit, where a fold mirror directs most of the beam (over a $113''  \times 113''$  FOV) into the main body of the imager, the MIRI Imaging Module (MIRIM)\footnote{A small fold mirror adjacent to the imager light path picks off the small FOV (up to a 8$''$  $\times$ 8$''$ FOV) of the spectrometer.}.   

The MIRIM optics are illustrated in Figure 2. The all-reflecting design was selected both because of the lack of suitable broad-band transmitting optical materials, and to assure achromaticity. For these reasons, all-reflecting optics are the norm for mid-infrared cameras and spectrometers \citep[e.g.][]{bryson1994, telesco1998, ettedgui-atad1998, kataza2000, packham2005}. MIRIM had the
additional challenge of a large format array and the resulting need for a large, aberration-free field of view. The adopted approach is
centered on a three-mirror anastigmat and is similar to the approach developed for the VLT Imager and Spectrometer for mid Infrared (VISIR) \citep{rio1998} and has also been used in the near infrared \citep[e.g.][]{mclean1998}. Mirror 1 (M1) is an ellipsoidal collimating mirror that forms a pupil where the cold stops and filters are placed. M2 is a flat folding mirror. M3, M4, and M5 are respectively hyperboloidal, ellipsoidal, and oblate ellipsoidal and together form the three mirror anastigmat, which images the input field onto the detector array at a net magnification of 0.352. That is, the telescope image scale has been compressed from 1.57 arcsec/mm to 4.4 arcsec/mm, giving a projected pixel scale of 0\farcs11. The FWHM of the JWST PSF at 5 $\mu$m is 0\farcs176; in other words, MIRIM Nyquist samples the PSF for wavelengths $\ge$ 6.25 $\mu$m, where the FWHM=0\farcs22. The optics also place the focal plane just outside the MIRIM housing to allow the focal plane module to be bolted directly to the housing in a very simple interface. The 1024 $\times$ 1024 pixel detector array intrinsic field is divided among pure imaging, low-resolution prism spectroscopy (Paper IV), and coronagraphy  (Paper V) as shown in Figure 1. Switching from one function to another entails pointing the telescope to place the source at the appropriate position within the overall focal plane and moving the filter wheel to place the appropriate optical element in the beam. 

A tolerance analysis of the MIRIM, including a budgeted wavefront error for the telescope, yields a Strehl of 92\% at 6 $\mu$m (assuming perfect focus between the telescope and the MIRIM entrance focal plane), and Strehls $>$ 80\% for defocus of up to $\pm$ 2mm. 
On this basis, it was concluded that the MIRIM would not require inflight focus adjustment; the instrument could be built to the budgeted tolerances.  The predicted values of distortion are $<$ 0.9\% over the entire field.

The primary cold stop for MIRI is placed at the telescope exit pupil formed near the fine steering mirror (FSM) outside the instrument and within the OTA, ahead of the MIRI POM. This pupil is imaged onto the front surface of the filters mounted in the filter wheel. A fixed cold stop is placed just in front of the filter wheel; this stop ensures there is no reflective path between the MIRIM entrance aperture and areas outside the edge of the filter wheel, where it could potentially leak to the detector without passing through a spectral filter. Masks are placed on each filter matched to the telescope pupil image to improve the rejection of thermal radiation from around the telescope primary. They are oversized to ensure no vignetting in the presence of small ($\Delta$R/D $\le$ 4\%) amounts of pupil shear. The motor that turns the wheel is the common version for MIRI described in Paper II. Its positioning accuracy of $\sim$ 1 arcsec places the individual stops to within 0.003\% of their diameters. The control of straylight is augmented by a cold stop around the imager entrance focal plane, i.e., blocking light incident outside the colored region in Figure 2. 

\subsection{Photometric Bands}

Filter positions are allocated as follows: 10 filters for imaging, 4 filter and diaphragm combinations for the coronagraphy (see Figure 3), 1 neutral density filter, 1 ZnS-Ge double prism for the LRS mode, 
1 opaque position for darks and 1 for a lens for ground test purposes. See Table\,\ref{Table 10} for a full listing. To obtain the relative response functions for the photometric bands, they must be multiplied by the optical transmission function (nearly neutral) and the detector spectral response (Rieke et al. 2014, Paper VII). Details on the resulting imager spectral bands are given in Table\,\ref{Table:filters}. All the  imager filters are broad band, and have been mainly chosen to observe continuum and silicate and aromatic features.

\subsection{Sub-arrays}

For MIRI observations 
of bright sources, faint sources near very bright ones, or high background fields, sub-arrays can be used (see the more detailed subarray description in Ressler et al. 2014, Paper VIII).  To ensure consistent calibration of SUBARRAY modes, only a few
sub-arrays will be made available and the user will have to select from these
predefined sets (Table 3). Tables 4 \& 5 give the current MIRI bright limits assuming use of the smallest practical subarray.

\section{Mechanical Design}

The MIRI imager consists of three main mechanical elements: the main structure, the three mirror anastigmat (TMA) assembly,
and the cover. The two other primary subsystems are the filter wheel and the focal plane module (FPM).
All except the cover and FPM are shown in Figure 4, which demonstrates how compactly the optical
system is folded. The entrance focal plane is to the left and can be identified by the coronagraph masks.
The light emerges to the right, through the square opening, where the FPM is installed, and the re-imaged focal plane falls on the 
detector sensitive area. Mirrors M1 and M2 are attached to the main structure from the
outside (as shown for M2). The internal alignment of the TMA camera is critical to its performance. By
building it as a separate unit, it can be aligned and then mounted onto the main structure with
relatively relaxed tolerances. The completed instrument is shown in Figure 5. 

To minimize differential thermal deformation, the primary and TMA structures and the cover of the MIRIM are fabricated 
of aluminium alloy 6061 T6. The largest components were machined from single aluminium blocks;
the units were stabilized by thermal cycling after rough machining was complete.  

The mirrors are of aluminium alloy 5000UP, because 
it can be polished to lower roughness ($<$ 10 nm rms) than the polishing
of 6061 alloy and thus better image quality and stray light levels can be achieved.
The optical surfaces of the mirrors are gold coated to improve the reflection of infrared light. 
A mirror and its support are made as a single piece, to eliminate any differential
deformations. The mirrors are attached to the structure by three elastic pads based on flexures, to avoid surface mirror
deformation due to the differential coefficients of thermal expansion after cool down to 6K or 
due to stress induced by the mounting screws.
To place the mirrors correctly, they are positioned by two pins 
and by the flatness of the mounting plane. Moreover, all mounting structure surfaces and all mirrors have a machined 
alignment reference surface that serves as a mirror. These surfaces allow positioning within tight tolerances by use of an autocollimator.

\section{Measured MIRIM performance}

\subsection{Optical Quality}

The predictions for the MIRIM optical performance are confirmed in practice. At a wavelength of 5.6 $\mu$m the image of a point source delivered by the JWST at the MIRI input focal plane should have a Full Width at Half Maximum (FWHM) of 0\farcs19, and the diameter of the first dark diffraction ring should be 0\farcs47 (these figures can be scaled to longer wavelength by multiplying by a factor $\lambda$/(5.6 $\mu$m)). Figure 6 shows the observed PSF using microstepping for maximum fidelity at CEA; we measure on this image FWHM = 0\farcs20 in the X direction and FWHM = 0\farcs19 in the Y direction (1.82 pix), matching the prediction for pure diffraction of 0\farcs19. Gaussian fits to the images measured in the Integrated Science Instrument Module (ISIM) at Goddard show that the focus is correctly placed; they yield a mean FWHM = 0\farcs21 at best focus, which is consistent with the diffraction-limited FWHM if allowance is made for the pixellation of the test images (0\farcs11 pixel width, and data taken without microstepping). In addition, these tests confirmed that the cold stop is aligned accurately within the ISIM.


The distortion in the MIRIM instrument has been measured at ambient temperature on the flight model at CEA-Saclay. The test consisted of imaging a grid of regularly spaced pinholes. The result is shown in Figure 7. Values range from 0\% (black) to 0.81\% (red) and are in very good agreement with the design predictions\footnote{The distortion is defined as the difference between the real position of the image and its ideal position if there were no distortion, the latter being computed considering the magnification in a given part of the field of view (the ``reference field", usually taken at the centre of the field of view). }.

\subsection{Out of Band radiation}

The broad spectral response of the MIRIM detector array makes blocking out-of-band (OOB) radiation a challenge. During RAL FM testing (Paper II) we did not detect significant OOB radiation that could compromise the instrument performance. This was verified using data taken with the MIRI Telescope Simulator (MTS) extended source at 800 K, the variable aperture system (VAS) fully open and at 2\% aperture, and the MIRI contamination control cover (CCC) open. We used a combination of the MTS blocking filters and MIRIM filters that should ideally end up with no signal detected. These observations were afterwards compared to exposures that used the exact same configuration but with the MTS blank in place. In the long wavelength range, light was dominated by a diffuse background component that was not affected by the presence of the MTS blocking filter. For this reason we could only state that the MTS MIRI filter combination is at least as effective as the MTS blank. However, for the short wavelength range the analysis proved that the blocking filter combination of MTS LWP and MIRI F560W is very efficient: only 0.1 to 0.2\% of the total light coming from the MTS was measured in the detector.

These results were afterwards confirmed during the first ISIM level testing campaign. In that case we observed the near-BB 100K emission of the Optical Telescope
Element Simulator (OSIM), and saw no evidence of spectral leaks with the F560W filter, at levels down to 1 part in 10$^6$.

\section{Operations}

\subsection{Overview}


The general MIRI operations concept is discussed in Gordon et al. (2014, Paper X). Here we focus on the MIRIM observations. 
The MIRIM data will have minimal on-board processing (by using loss-less data compression and an appropriate 
rearrangement of the image, see Paper VIII). Some high-volume data cases will also use coaddition to manage the data volume. 
The full array mode will be used for nearly all imaging. The images are very well sampled; even with the shortest wavelength band at 5.6 $\mu$m, the pixels are within $\sim$ 30\% of nominal Nyquist sampling. Finer sampling can be obtained through intensive dithering, supporting such applications as super-resolution. Observer-selected parameters for imaging include filter selection, integration time(s), and telescope pointing (dithering/mapping pattern). Occasionally, use 
of sub-arrays may be warranted, which will set the read out mode. In full-array mode, normal read-out 
will result in a sampling interval of 2.775 sec. 
  
The full suite of components mounted in the MIRIM 18-position filter wheel is listed in Table\,\ref{Table 10}. 
The instrument control electronics (ICE) moves the filter wheel when 
commanded by the MIRI flight software (FSW). The ICE can measure the absolute position of the filter wheel 
before and/or after a movement; (e.g. the filter wheel is at position 7), and can move it 
to the next adjacent position in either direction. The filter is located with a ratchet system utilizing a spring loaded detent arm.

The generic readout pattern used by the MIRIM sensor chip assembly is described elsewhere 
(Paper VIII). In practice, to set the total time integrating on
a source the integration time selected by the observer is specified as the number of groups in an integration (where a group is the set of data resulting
from cycling through all of the pixels in the array) and the number of integrations in an exposure. 
The term ``group" is interchangeable with ``frame". Another possible variable is the number of 
samples per pixel while the pixel is addressed, but this
parameter will be set automatically according to the type of observation. For MIRIM the 
array readout will generally be in the ``FAST" mode (``SLOW'' is reserved for spectroscopy) and a pixel will be sampled once. The quantum of 
time for reading out a MIRI array is then 2.775 seconds, the minimum time to cycle through all the pixels. 
The integration time will then approximately be the number of groups selected times 2.775 seconds, e.g. if
10 groups are selected, they will yield an integration of 26.51 seconds. This pattern is repeated
in an observation as necessary to yield the desired total integration time.  

The high sensitivity of the MIRI imager will tempt obervers to plan complex programs of short integrations. However, realistic programs can also be subject to significant time overheads. 
Some examples of the current best estimates for the overhead durations are 
given in Table\,\ref{Table 11}.

\subsection{Dithering}

The optimum observing strategy for point or slightly extended sources is to dither by a small angle
every five minutes or so. This approach allows building a sky flat that is free of artifacts due
to slow drifts inherent in the detector array and its readout. Similar observing strategies were
found to be best with the similar detector arrays used in the three {\it Spitzer} instruments.  
It is recommended that all appropriate observations include dithers, in part
so the data archive will have uniform products.   
Dithering will require moving the telescope; relatively small moves (probably $<$ 30$''$) can be made without
reacquiring guide stars and suffering the large resulting overhead  (Table\,\ref{Table 11}). Observers will be
offered a menu of options selected for high efficiency and to allow removal of bad pixels and 
improve the sampling of the point spread function (PSF). 
The imager dithering is discussed further in Paper X. 

Such small angle dithering is not optimum for sources extended by more than $\sim$ 10$''$. 
Approaches are under development for the imaging and mapping of extended sources.

\section{Calibration}

Calibration of the MIRIM will be done periodically 
to monitor its stability and performance and will thus not be the observer's responsibility. Diagnostic tests and 
maintenance will also be done as needed. The design of the Calibration System, discussed in Paper II, is driven by the need to correct the
various detector artifacts (Papers VII and VIII), and then to carry out the usual steps of 
subtracting the background signal level from an astronomical observation, and correcting for variations in response across the resulting images due to pixel 
to pixel gain variations and changes in optical transmission across the field. 

\subsection{Imager Pixel Flat Field}

The MIRIM pixel flat field (i.e. detector gain matrix) has been based on measurements obtained during the Flight Model (FM) testing campaign carried out at RAL (Rutherford Appleton Laboratory) during spring-summer 2011. All data used the MIRI Telescope Simulator (MTS) extended source (Paper II). 

The pixel flats for both FULL frame readout modes, FAST and SLOW, and of the BRIGHTSKY and SUB256 subarrays were studied (see Table 3, and Papers VIII and X for a subarray description). The SLOW and subarray modes were used only to compare with the FULL FAST flat field in one particular filter, and test if it is necessary to have dedicated flat fields for each detector readout mode. 
Raw data were reduced with the MIRI DHAS (Data Handling and Analysis System), that provides the slope image in DN/s. After standard processing, all FULL frame data were clipped using the median value of a 220$\times$180 pix$^2$ clean area of the detector. Pixels with values below 0.5$\times$median and above 1.5$\times$median were set to NaNs. Pixels located in the four quadrant mask (4QPM) coronagraphic area of the detector were set to NaN-values as well. This way we excluded the dark areas and hot/dead pixels. Large scale inhomegeneities were afterwards removed by smoothing the clipped images using a 10$\times$10 pix$^{2}$ box. We then obtained the pixel flat by dividing the clipped image by the clipped and smoothed one. When several observations were available, the individual flat fields were combined into a median one.
The same analysis was performed in the subarrays, only changing the area used to define the clipping.

The MIRIM detector exhibits a difference between odd and even rows (Paper VIII). This difference, however, appears to be dependent on the detector illumination level, and as such should not be present in the pixel flat. It was hence calculated by subtracting one from the median of odd and even rows, and removing this offset from the data of even and odd rows, respectively. Figure\,\ref{fig:flat-even-odd} shows the different distributions between odd and even rows for the F1300W filter, and the final result with the rows difference successfully removed.

It is recommended to use wavelength dependent flat fields to remove from the data spurious circular patterns that slightly change with wavelength, and that are successfully eliminated by the pixel flat. The comparison of the distributions of FULL array FAST readout flat field for the F1130W filter with the SLOW mode and subarrays indicates that dedicated flat fields should be used for each readout mode/filter.

\subsection{Dark Current Measurement}

The blank position on the filter wheel allows measurement of  the dark current. Those measurements are discussed in Paper VIII. 
The detector dark current is around 0.2 el/sec/pixel, and its
contribution to an integration will normally be removed by taking the difference of dithered images. 
 MIRIM has regions which do not see astronomical signal during normal operations. 
These regions (e.g., behind the frame holding the coronagraphic masks), will allow dark current monitoring of all MIRI frames.
A case for taking additional very high signal to noise ratio dark current measurements may arise if such
monitoring shows that the dark 
current is also very stable, such that it changes by less than the read noise on long timescales. 
This ``super-dark" measurement would be particularly useful for LRS observations taken at short wavelengths where the 
dark current is a significant part of the total signal. If it were subtracted from the raw data, instead of using the normal off-source observation, 
it could provide a gain in the signal to noise at little expense, since the super-dark could be taken in parallel with other (MRS) 
observations and hence at no cost to observing efficiency. 

The MIRIM can also make periodic measurements of 
the dark current in all pixels.
This ability to occasionally measure a dedicated dark image may be useful for subtraction 
from both the sky and on-board flat field
observations as part of the measurement of the detector gain matrix. 
A dark image can be obtained either by using the filter wheel blank position or by 
closing the Contamination Control Cover. In the latter case, it is possible also to operate the 
Calibration Source (Paper II) to allow a measurement of its flux alone, without the contribution of sky 
and telescope backgrounds.

\subsection{Spectrophotometric Performance}

The MIRIM spectrophotometric performance has been studied using data from the FM testing campaign at RAL. 
The relative flux calibration factors were estimated using the MTS extended source; the use of point source data was ruled out due to discrepancies between measured and expected signals as predicted by MTSSim, (MTS Simulator, Paper II).  All extended source data were taken using the MTS black body at 200K, the FULL array in  FAST readout mode, and the VAS at 100\% (i.e. fully open). Exposure times were chosen for the detector to reach half-well (about 2.5$\times$10$^4$ DN). Complementary background data were taken for each integration by placing the MTS filter in the blank position. 
Raw data were processed with the MIRI DHAS, including cosmic ray removal, background subtraction and filter-dependent flat field correction. 

Calibration factors were estimated by comparing the measured slopes with the simulated flux density from MTSSim converted to an object on the sky in Jy/arcsec$^2$ (see Table 7). For this particular case in which the source SED is known, colour corrections were calculated and applied.

The MIRIM absolute calibration will not be fully determined until the telescope is on-orbit. At that time a set of stars (A0V stars, solar-type stars, and white dwarfs) preselected to have accurately predicted fluxes in all bands will be measured and will determine the final conversion. Currently, results are limited by the accuracy of the knowledge of the MIRI Telescope Simulator and of the transmissions of the MIRI sub-systems (Paper IX). However, the results are in good agreement with expectations from the properties of the instrument components, giving confidence that the instrument optics is working at high efficiency.

The conversion between number of incident photons at the MIRI entrance focal plane and number of measured electrons (Photo Conversion Efficiency, PCE) for each MIRIM filter has also been derived. The estimated PCE is a relative one, as it has been based on the MTS emission predicted by MTSSim (see also Paper IX). The measured slope values at the Focal Plane Array (FPA) were converted to electron flux at the MIRI entrance plane, using a gain value of 5.5$e^{-}$/DN (Paper VIII), the pixel area and the filter widths. These measurements were afterwards compared to the MTSSim predicted emission at the entrance focal plane in photons allowing us thus to estimate the relative PCE for each MIRIM filter (see Table 7, see also Paper IX).

\subsection{Spatial Photometric Repeatability}

Consistency between point source measurements throughout the detector plane has also been verified using RAL FM data. The 100 and 25\,$\mu$m MTS pinholes (Paper II) were used to scan the imager part of the detector using  9 locations with each point source and in all filters (see Fig. 9). All integrations used the MTS BB at 800~K. Exposure times and VAS positions were chosen to reach detector half-well and prevent saturation (the 800~K emission is very strong at short wavelengths).

Data reduction included cosmic ray removal, background subtraction and filter-dependent flat field and linearity correction.  
Aperture photometry was performed on all 100\,$\mu$m and 25\,$\mu$m point source data, using an annulus to measure and subtract the background. The estimated noise was based on the rms sky annulus.
This may not be optimal for absolute calibration purposes, but since only relative differences between the point sources in the same filter are of interest here, this choice is justified. An annulus with inner and outer radius of 45 and 50 pixels, respectively, was used in all cases to subtract the background.  

The measured standard deviation over the point source fluxes at the different positions ranges from 9.4 to 24.8\% for filters F560W to F2550W, respectively. We explain these large deviations being partially due to vignetting caused by the mechanical structure of the MTS pinhole (monochromatic model by M. Wells, priv. comm.), but mostly because of wavelength dependent effects resulting from the individual structures of the pinholes themselves. This hypothesis is confirmed by analyzing the differences between the fitted surfaces to the 25\,$\mu$m and 100\,$\mu$m point source observations over the entire detector. After applying the monochromatic model and correcting with a surface model, the 100\,$\mu$m pinhole shows a standard deviation of less than 2\%. This is not the case for the 25\,$\mu$m pinhole, where the non-ideal effects are largest, as a result of differences between the fitted surfaces. These results indicate that the measured (uncorrected) standard deviation is related to the construction of the two pinholes as part of the MTS, rather than being intrinsic to the MIRI imager detector. This conclusion indicates that we can expect good in$-$flight stability and spatial uniformity.

\subsection{Cross calibration with the Medium Resolution Spectrometer}
One important aspect of the spectrophometric performance of MIRI at RAL has been the verification of the cross-calibration between the MIRIM and the MRS. We based the cross-calibration on 100\,$\mu$m point source data in both imager and spectrometer. For MIRIM all filters were used with the MTS BB temperature set at 800\,K and a fully open VAS. To prevent detector saturation at the shortest wavelengths (filters F560W to F1500W) we used the AXIS64 subarray, that was specifically defined at RAL for testing purposes. The AXIS64 is centred at [518, 514] in the FULL frame, and presents an effective illuminated area of 64$\times$64 pix$^{2}$. Longer wavelength filters were used with FULL frame readout. 

After standard data reduction, aperture photometry was performed on the images, and the relative calibration factors derived as explained above were used to calibrate the point source observations. Because of the change in observed flux level depending on the position of the point source on the imager detector (as discussed above), a single achromatic correction factor was applied to account for the vignetting of the 100 $\mu$m pinhole (77.41\%, based on the geometric model provided by M. Wells, priv. comm.) and thus to correct the results for this effect before comparing them with the MRS ones.

The short- (SW) and long-wavelength (LW) MRS detectors were also used to observe the 100 $\mu$m point source with the exact MTS configuration used for the imager. The spectral cube was derived and afterwards calibrated relatively with respect to the MTS following \citet{Wel14}, Paper VI. As before, MRS spectra were also corrected for vignetting using the achromatic model. The calibrated spectrum was extracted from the spectral cube by applying aperture photometry on each spectral element in the cube. Allowance was made for the different spaxel sizes in the MRS bands  to extract equivalent regions over the full spectral range.

At the shortest wavelengths, the match between the MRS and the imager calibration differs by only 8.4\% at 5.6$\mu$m, and 1.5\% at 7.7\,$\mu$m. Towards longer wavelengths we see an increasing discrepancy between the imager and the MRS, which we attribute to the fact that no wavelength dependent corrections were applied, e.g. for the reflectance inside the vignetting ’tunnel’ obscuring the 100\,$\mu$m pinhole. We thus conclude that the discrepancy at longer wavelengths can be attributed to the test setup and not to the MIRI.

\section{Acknowledgments}
The work presented is the effort of the entire MIRI team and the enthusiasm within the MIRI partnership is a significant factor in its success. MIRI draws on the scientific and technical expertise many organizations, as summarized in Papers I and II. 
A portion of this work was carried out at the Jet Propulsion Laboratory, California Institute of Technology, under a contract with the National Aeronautics and Space Administration.

We would like to thank the following National and International
Funding Agencies for their support of the MIRI development: NASA; ESA;
Belgian Science Policy Office; Centre Nationale D'Etudes Spatiales;
Danish National Space Centre; Deutsches Zentrum fur Luft-und Raumfahrt
(DLR); Enterprise Ireland; Ministerio De Econom\'ia y Competividad;
Netherlands Research School for Astronomy (NOVA); Netherlands
Organisation for Scientific Research (NWO); Science and Technology Facilities
Council; Swiss Space Office; Swedish National Space Board; UK Space
Agency.

\clearpage

\begin{deluxetable}{lcccl}
\tabletypesize{\scriptsize}

\tablecaption{List of Components Mounted in the MIRIM Filter Wheel \label{Table 10}}
\tablewidth{0pt}
\tablehead{

\colhead{Filter Name} & \colhead{$\lambda_0$ [$\mu$m]} & \colhead{$\Delta\lambda$ [$\mu$m]} & pos. 
& \colhead{Comment}}

\startdata 
F560W  &  5.6 & 1.2 &  10 &Broad Band Imaging \\  
F770W  &  7.7 & 2.2 & 13 & PAH, broad band imaging \\
F1000W & 10.0 & 2.0 &  3 & Silicate, broad band imaging \\  
F1130W & 11.3 & 0.7 &  4 & PAH, broad band imaging \\ 
F1280W & 12.8 & 2.4 & 5 & Broad band imaging \\
F1500W & 15.0 & 3.0 & 7 & Broad band imaging \\ 
F1800W & 18.0 & 3.0 & 8 & Silicate, Broad band imaging \\ 
F2100W & 21.0 & 5.0 & 9 & Broad band imaging \\ 
F2550W & 25.5 & 4.0 & 15 & Broad band imaging \\ 
F1065C & 10.65 & 0.53 & 18 & Phase mask, NH3, silicate \\ 
F1140C & 11.40 & 0.57 & 16 & Phase mask, continuum or PAH feature \\  
F1550C & 15.50 & 0.78 & 14 & Phase mask, continuum \\ 
F2300C & 23.0 & 4.6 & 12 & Focal plane mask, peak of debris \\ 
F2550WR & 25.5 & 4.0 & 17 & Redundant imaging filter, risk reduction \\ 
FND & $\sim$ 13 &  $\sim$ 10 & 1 & $\sim 2 \times 10^{-3}$ ND for coronagraphy acquisition \\ 
FLENS & N/A & N/A & 11 & Testing \\ 
P750L & N/A & N/A & 6 & Prism for Low Resolution Spectroscopy \\ 
OPAQUE & Blackened blank & N/A & 2 & Dark measurements \\ 
\enddata

\end{deluxetable} 

\clearpage

\begin{deluxetable}{lccc}
\tabletypesize{\scriptsize}
\tablecaption{MIRIM Image Sizes and Sensitivities.\label{Table:filters}}
\tablewidth{0pt} 
\tablehead{
\colhead{Filter} & \colhead{FWHM [arcsec]} & 
\colhead{80\% Encircled Energy} & \colhead{10$\sigma$;10,000sec} \\ 
\colhead{} & \colhead{} & \colhead{Diameter [arcsec]} & \colhead{sensitivity [$\mu$Jy]}}

\startdata
F560W        & 0.22 & 0.61    &  0.17 \\
F770W      &    0.25 & 0.85    &  0.27 \\ 
F1000W      &  0.32 & 1.10    &  0.60 \\
F1130W      &  0.36 & 1.24    &  1.48 \\
F1280W     &  0.41 & 1.41    &   0.94 - 1.05* \\
F1500W     &  0.48 & 1.65    &   1.5 - 2.0 \\
F1800W     &  0.58 & 1.98    &   3.7 - 5.3 \\
F2100W     &  0.67 & 2.31    &   7.5 - 10.5 \\
F2555W/WR  & 0.82 & 2.81    &   27 - 36 \\
\enddata
\tablenotetext{*}{Ranges of values reflect uncertainty in the emission level from the JWST observatory.}
\end{deluxetable} 

\clearpage

\begin{deluxetable}{lrrrcc}
\tabletypesize{\scriptsize}
\tablecaption{MIRI Imager Sub-array Locations and sizes\label{Table 8}}
\tablewidth{0pt}
\tablehead{
\colhead{Subarray} & \colhead{Rows} & \colhead{Columns} & \colhead{First Row} & 
\colhead{First Col.} & \colhead{Light Sensitive Columns}}

\startdata
FULL         & 1024 & 1032 & 1 & 1    &   1024 \\
MASK1065      & 224 & 288 &   19 & 1    &   224 \\ 
MASK1140      & 224 & 288 & 245 & 1    &   224 \\
MASK1550      & 224 & 288 & 467 & 1    &   224 \\ 
MASKLYOT      & 304 & 320 & 717 & 1    &   275 \\
BRIGHTSKY     & 512 & 512 &   51 & 457    &   512 \\
SUB256        & 256 & 256 &   51 & 413    &   256 \\
SUB128        & 128 & 136 & 889 & 1    &   128 \\
SUB64         &  64 & 72  & 779 & 1    &    64 \\
SLITLESSPRISM & 416 & 72  & 529 & 1    &    64 \\
\enddata
\end{deluxetable}

\clearpage

\begin{deluxetable}{lccccc}
\tabletypesize{\scriptsize}
\tablecaption{MIRI Imager Sub-array Frame Times and Maximum Fluxes (60\% full well)\label{Table 9}}
\tablewidth{0pt}
\tablehead{
\colhead{Subarray} & \colhead{Size Columns by Rows} & \colhead{Start Pos} & \colhead{FAST Frame Time (s)} & 
\colhead{Max Flux F560W [mJy]} & \colhead{Max Flux F2550W [mJy]}}
\startdata
FULL       &   $1032 \times 1024$  & (1,1) & 2.78 &  13 &  202 \\
BRIGHTSKY  & $512\times512$ & (457,51) & 0.865 &  42 &  540 \\
SUB256     & $256\times256$ & (413,51) & 0.300 & 120 & 2000 \\
SUB128     & $136\times128$ & (1,899) & 0.119 & 300 & 4700 \\
SUB64      &  $72\times64$  & (1,779) & 0.085 & 420 & 6600 \\
\enddata
\end{deluxetable} 

\clearpage

\begin{deluxetable}{lcc}
\tabletypesize{\scriptsize}
\tablecaption{Bright Limits (Vega magnitudes) in imaging mode with the 64 x 64 subarray \label{Table 7}}
\tablewidth{0pt}
\tablehead{
\colhead{Filter Name} & \colhead{G0 star (K mag)} & \colhead{M5 star (K mag)}}

\startdata
F560W     & 6.2  &  6.7 \\
F770W      & 6.1 & 6.6 \\ 
F1130W      & 2.8 & 3.4 \\
F1500W     & 2.9 & 3.5 \\
F2100W    & 1.5  &  2.1 \\
\enddata
\end{deluxetable} 

\clearpage 

\begin{deluxetable}{lc}
\tabletypesize{\scriptsize}
\tablecaption{Overhead examples used in Calculating MIRIM Efficiency.\label{Table 11}}
\tablewidth{0pt}
\tablehead{
\colhead{MIRIM Activity} & \colhead{Overhead [seconds]}} 
\startdata
Large angle slew to target  &  1800 \\
Acquire initial guide stars  &  240 \\
Average time to reach any filter (6 steps) & 60 \\ 
Initialize detector array  &   5  \\
Dither ($\lesssim$ 30$''$) & $\sim$ 20 - 40  \\
Larger dither (\& re-acquire guide star)  &  300 \\
Time to close CCC & 60 \\ 
Time to turn on the calibration lamp & 60 \\ 
Time to re-acquire guide stars (for large dithers) & 240 \\
\enddata
\end{deluxetable}

\clearpage

\begin{deluxetable}{lcc}
\tabletypesize{\scriptsize}
\tablecaption{Relative calibration parameters for all MIRIM filters.\label{Table 12}}
\tablewidth{0pt}
\tablehead{
\colhead{Filter}  &  \colhead{PCE}  & \colhead{Calibration Factor$^1$} \\    
\colhead{}  & \colhead{e$^-$/photon}  & \colhead{Jy/arcsec$^2$ per DN/s}
}
\startdata
    F560W   & 0.3157  $\pm$ 0.120   &  $2.41 \times 10^{-5}$  \\ 
    F770W   & 0.3210 $\pm$ 0.0835  &  $1.32 \times 10^{-5}$  \\ 
    F1000W  & 0.3685 $\pm$ 0.0923  &  $1.76 \times  10^{-5}$ \\
    F1065C  & 0.2010 $\pm$ 0.0405  &  $1.37 \times 10^{-4}$  \\
    F1130W  & 0.3857 $\pm$ 0.0801  &  $5.76 \times 10^{-5}$  \\
    F1140C  & 0.1880 $\pm$ 0.0371  &  $1.43 \times 10^{-4}$  \\
    F1280W  & 0.3443 $\pm$ 0.0769  &  $2.11 \times 10^{-5}$  \\
    F1500W  & 0.3498 $\pm$ 0.0715  &  $1.84 \times 10^{-5}$  \\
    F1550C  & 0.1488 $\pm$ 0.0283  &  $1.81 \times 10^{-4}$  \\
    F1800W  & 0.3115 $\pm$ 0.0617  &  $2.68 \times 10^{-5}$  \\
    F2100W  & 0.2885 $\pm$ 0.0517  &  $2.04 \times 10^{-5}$  \\
    F2300C  & 0.1810 $\pm$ 0.0323  &  $3.65 \times 10^{-5}$  \\
    F2550W  & 0.2112 $\pm$ 0.0380  &  $4.42 \times 10^{-5}$  \\
    F2550WR & 0.1894 $\pm$ 0.0340  &  $ 4.60 \times 10^{-5}$  \\
   \hline
\enddata
\tablenotetext{1}{Values are per pixel.}
\end{deluxetable}

    

\clearpage	

\clearpage

\begin{figure}[t] 
\begin{center}  
\includegraphics[width=6.0in]{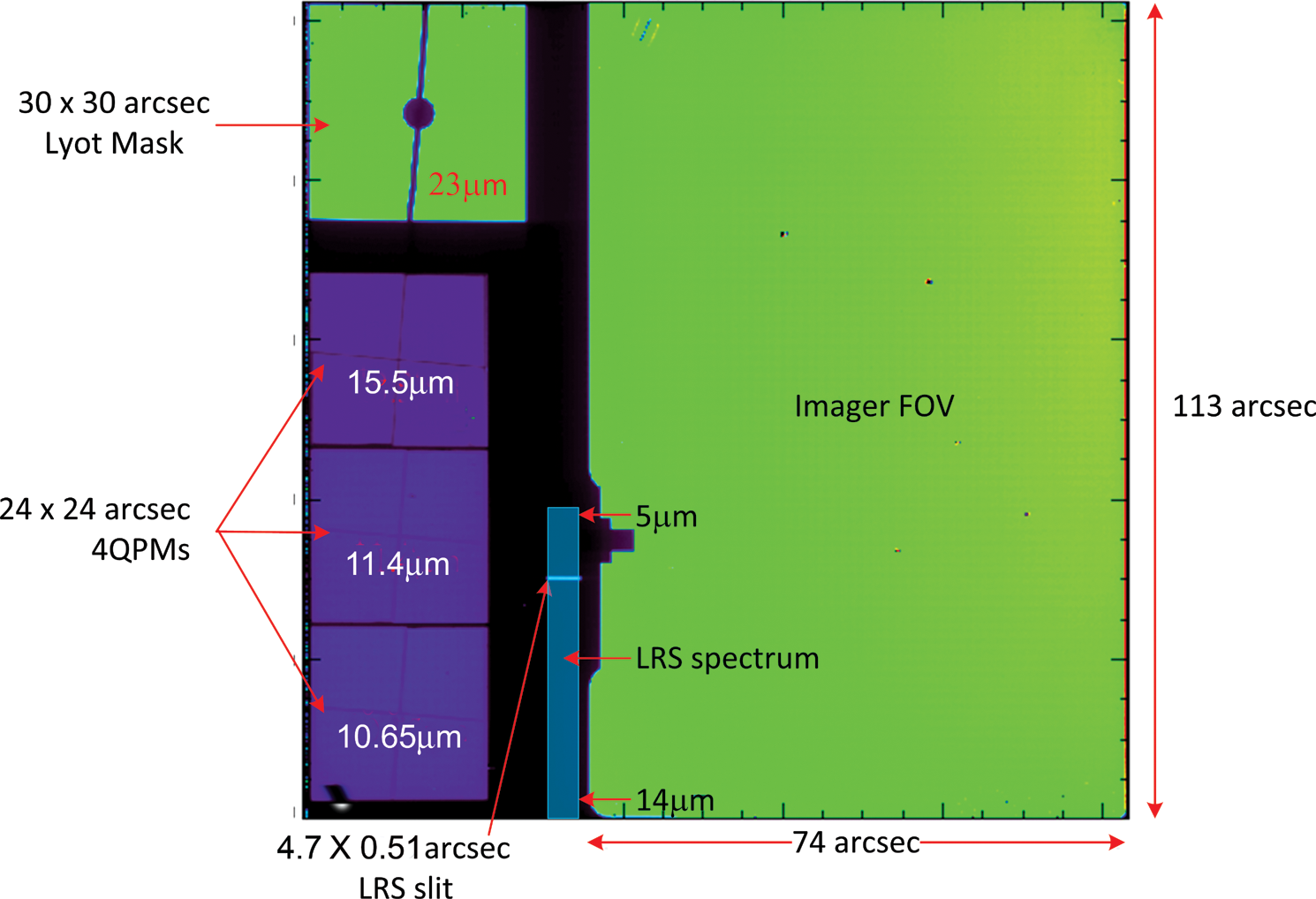}  
\caption{MIRIM input focal plane. The green area to the right is the clear imager section of the array. The three 4-quadrant-phase-mask coronagraphs and the Lyot coronograph are located along the left edge; the latter has an occulting spot 2\farcs16 in radius. 
The Low Resolution Spectrometer slit  location is also shown, along with the wavelength range of the spectra, which provide a typical resolution of R $\sim$ 100  from 5 to 12 $\mu$m (with some response to 14 $\mu$m).} \label{fig:mirim-focal-plane}  
\end{center}  
\end{figure}  

\clearpage

\begin{figure}[t] 
\begin{center}  
\includegraphics[width=6.0in]{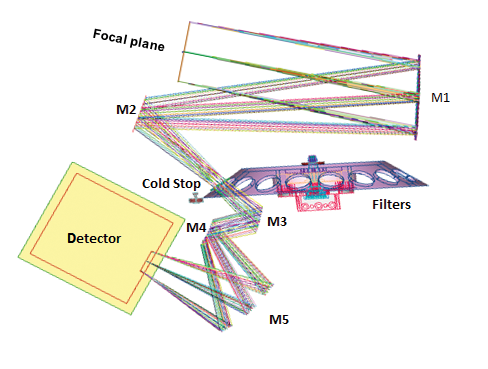}  
\caption{The MIRIM optical layout. M1 forms the pupil where the filters and cold stops are placed, M2 folds the beam, and M3 - M5 form an anastigmat that reimages the telescope focal plane onto the detector array} \label{fig:optical-layout}  
\end{center}  
\end{figure}  

\clearpage

\begin{figure}[h] 
\begin{center}  
\includegraphics[width=5.0in]{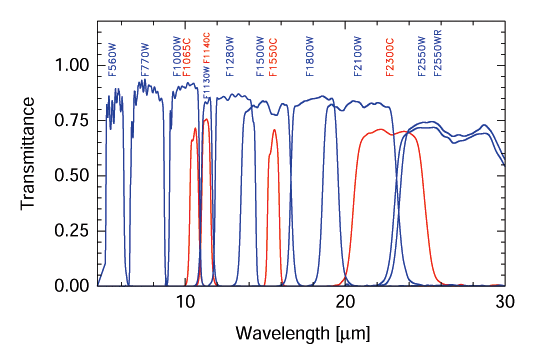}  
\caption{ MIRIM filter transmission profiles. The imager filters are depicted in blue, and the coronograph ones in red.} \label{fig:filters}  
\end{center}  
\end{figure}  

\clearpage

\begin{figure}[h] 
\begin{center}  
\includegraphics[width=7.0in]{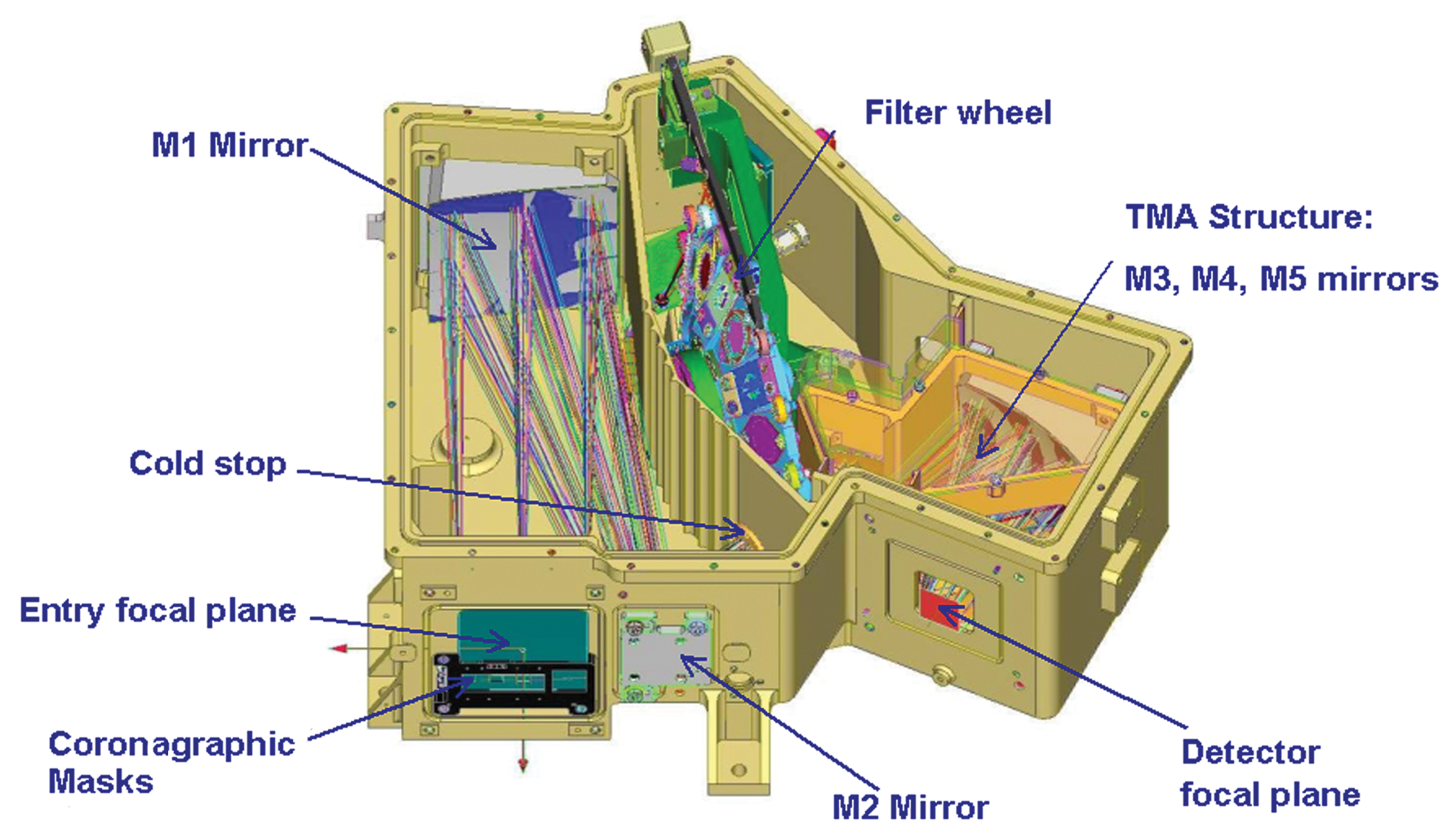}  
\caption{Mechanical layout of the MIRIM. The optical design permits folding into a
very compact, rigid structure. The focal plane module is not shown but is attached in the middle right in this illustration (from http://smsc.cnes.fr/MIRI/).}  \label{fig:filters}  
\end{center}  
\end{figure}  

\clearpage

\begin{figure}
\begin{center}  
\epsscale{1.0}
\includegraphics[width=5.0in]{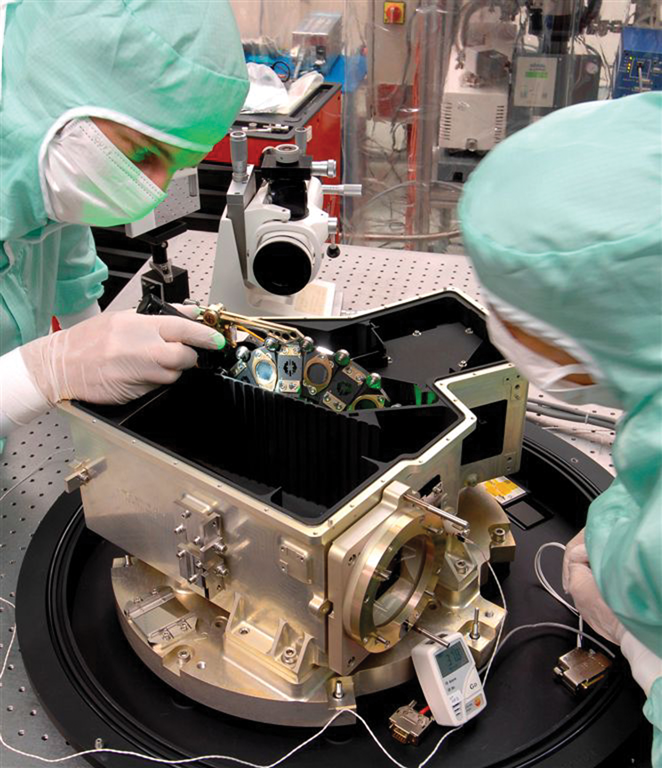}
\caption{ The MIRIM under a late-stage inspection prior to integration with the rest of the instrument (from http://smsc.cnes.fr/MIRI/). 
The instrument is rotated about 60$^o$ counterclockwise compared with the preceding figure. \label{fig:picture}}
\end{center}  
\end{figure}

\clearpage

\begin{figure}
\begin{center}  
\epsscale{1.0}
\includegraphics[width=7.0in]{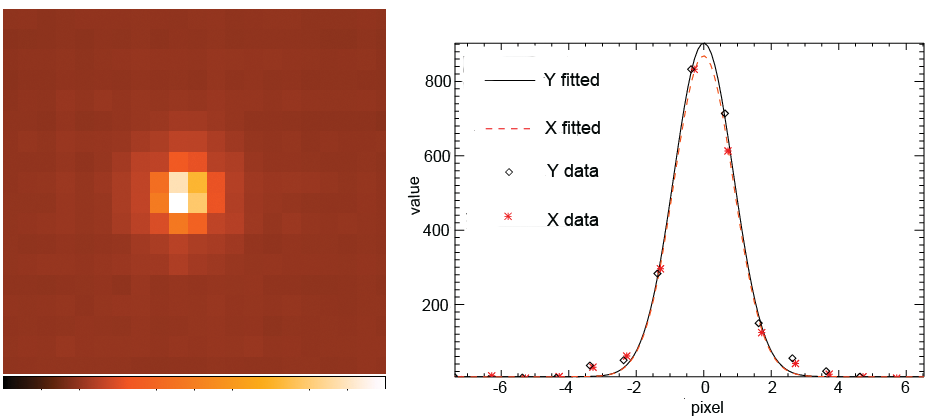}
\caption{ The MIRIM PSF as observed at GSFC in October 2013 through the F560W filter (left), and slices in the X (red stars) and Y (black empty diamonds) directions fitted by Gaussian curves\label{fig:PSF-ISIM}}
\end{center}  
\end{figure}

\clearpage

\begin{figure}
\begin{center}  
\epsscale{0.8}
\includegraphics[width=4.0in]{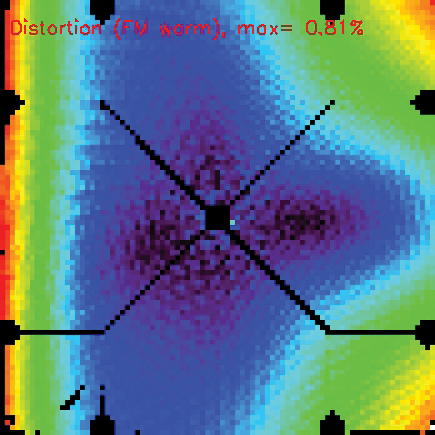}
\caption{ The distortion map in the MIRIM (black is 0\%, red is 0.81\%; see text). The black
lines and sharp-edged areas are artifacts of the test apparatus. \label{fig:distorsion}}
\end{center}  
\end{figure}

\clearpage

\begin{figure}  
\begin{center}  
\includegraphics[width=5.0in]{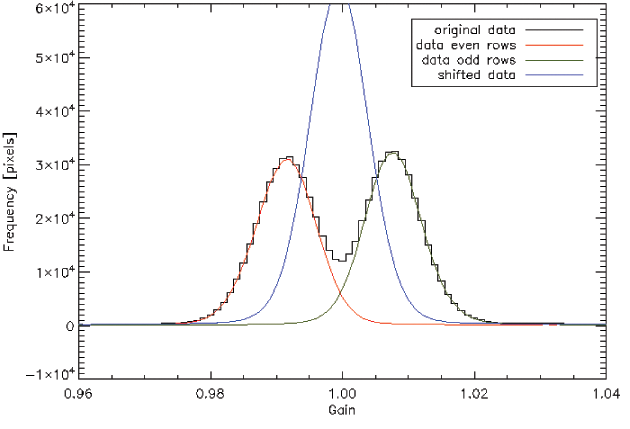}  
\caption{Pixel distribution for the 800\,K dataset, filter F1130W. The black line represents the histogram for all pixels, whereas red and green show the odd and even row pixels, respectively. The blue line represents the distribution of all the pixels after the successful removal of the odd-even row effect.} \label{fig:flat-even-odd}  
\end{center}  
\end{figure}

\clearpage

\begin{figure}  
\begin{center}  
\includegraphics[width=5.0in]{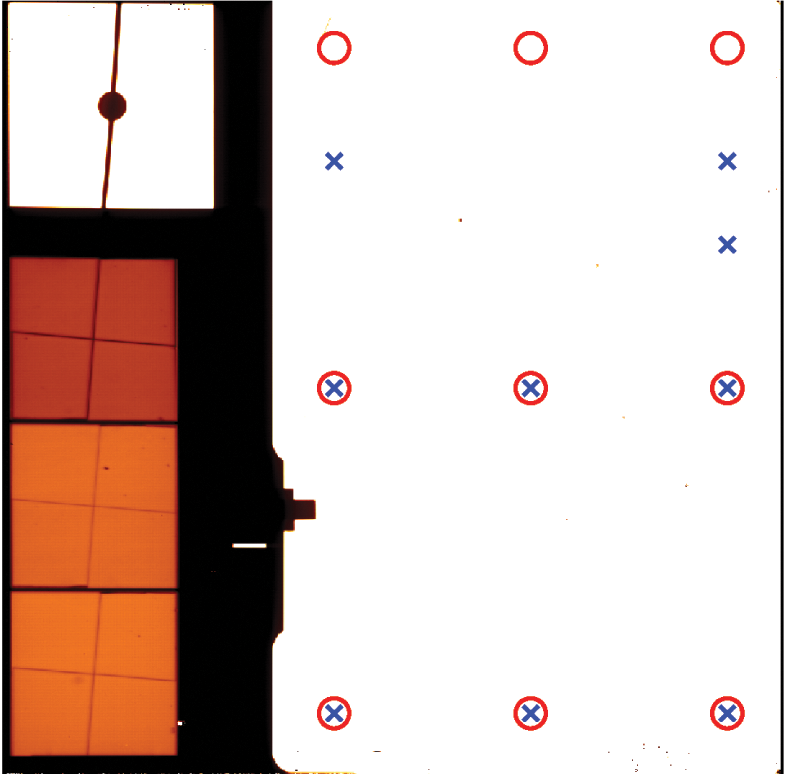}  
\caption{Point source positions distributed over the detector used to verify the photometric repeatability. Circles and crosses indicate the locations of the 100 and 25\,$\mu$m point sources, respectively.} \label{Figure 8}  
\end{center}  
\end{figure}

\end{document}